%% file: main.tex
\documentclass[10pt,conference]{IEEEtran}
\IEEEoverridecommandlockouts
\usepackage{cite}
\usepackage{amsmath,amssymb,amsfonts}
\usepackage{algorithmic}
\usepackage{graphicx}
\usepackage{textcomp}
\usepackage{xcolor}
\usepackage{color,soul}
\usepackage{graphicx, caption, subcaption}
\def\BibTeX{{\rm B\kern-.05em{\sc i\kern-.025em b}\kern-.08em
    T\kern-.1667em\lower.7ex\hbox{E}\kern-.125emX}}

\begin{document}

\title{Collective Communication Patterns Using Time-Reversal Terahertz Links at the Chip Scale\\

\thanks{\textsuperscript{*}F. Rodríguez-Galán and A. Bandara contributed equally to this work.
\newline Authors acknowledge support from the European Union’s Horizon 2020 research and innovation program, grant agreement 863337 (WiPLASH), the European Research Council (ERC) under grant agreement 101042080 (WINC) as well as the European Innovation Council (EIC) PATHFINDER scheme, grant agreement No 101099697 (QUADRATURE). Corresponding author: Sergi Abadal (abadal@ac.upc.edu)}
}

\author{\IEEEauthorblockN{F\'atima Rodr\'iguez-Gal\'an*}
\IEEEauthorblockA{\textit{NaNoNetworking Center in Catalunya}\\
\textit{Universitat Polit\`{e}cnica de Catalunya}\\
Barcelona, Spain \\
}
\and
\IEEEauthorblockN{Ama Bandara*}
\IEEEauthorblockA{\textit{NaNoNetworking Center in Catalunya}\\
\textit{Universitat Polit\`{e}cnica de Catalunya}\\
Barcelona, Spain \\
}
\and
\IEEEauthorblockN{Elana Pereira de Santana}
\IEEEauthorblockA{\textit{Institute of High Frequency and Quantum}\\
\textit{Electronics, University of Siegen}\\
Siegen, Germany}
\and
\IEEEauthorblockN{Peter Haring Bol\'ivar}
\IEEEauthorblockA{\textit{Institute of High Frequency and Quantum}\\
\textit{Electronics, University of Siegen}\\
Siegen, Germany}
\and
\IEEEauthorblockN{Eduard Alarc\'on}
\IEEEauthorblockA{\textit{NaNoNetworking Center in Catalunya}\\
\textit{Universitat Polit\`{e}cnica de Catalunya}\\
Barcelona, Spain}\\
\and
\IEEEauthorblockN{Sergi Abadal}
\IEEEauthorblockA{\textit{NaNoNetworking Center in Catalunya}\\
\textit{Universitat Polit\`{e}cnica de Catalunya}\\
Barcelona, Spain}\\

}

\maketitle

\begin{abstract}
Wireless communications in the terahertz band have been recently proposed as complement to conventional wired interconnects within computing packages. Such environments are typically highly reverberant, hence showing long channel impulse responses and severely limiting the achievable rates. Fortunately, this communications scenario is \emph{static} and can be pre-characterized, which opens the door to techniques such as time reversal. Time reversal acts a spatial matched filter and has a spatiotemporal focusing effect, which allows not only to increase the achievable symbol rates, but also to create multiple spatial channels. In this paper, the multi-user capability of time reversal is explored in the context of wireless communications in the terahertz band within a computing package. Full-wave simulations are carried out to validate the approach, whereas modulation streams are simulated to evaluate the error rate as a function of the transmitted power, symbol rate, and number of simultaneous transmissions.  
\end{abstract}


\section{Introduction}
\input{1-intro-short}

\section{Background}
\label{sec:background}
\input{2-background-short}

\section{System Model} 
\label{sec:model}
\input{3-model}

\section{Performance Evaluation of Multi-Channel Time-Reversal within Package}
\label{sec:results}

\input{4-results-short}

\section{Discussion}
\label{sec:discussion}
\input{5-discussion-short}

\section{Conclusion}
\label{sec:conclusion}
\input{6-conclusion-short}

\bibliographystyle{IEEEtran}
\bibliography{IEEEabrv,bib}


\end{document}

%% file: 1-intro-short.tex
Terahertz communications open the door to a wealth of opportunities in 6G networks and beyond due to not only the much-sought increase in bandwidth, but also the miniaturized size of the required antennas and transceivers \cite{lemic2021survey}. Thanks to such miniaturization, new applications arise as communications can be taken to inaccessible locations such as inside the human body \cite{saeed2021body} or within computing packages both in the classical \cite{chen2019channel, Yi2021} and quantum computing domains \cite{alarcon2023scalable}.

Being considerably different than traditional wireless communication scenarios, novel applications often pose new challenges in the form of tight resource constraints (e.g. energy from harvesting sources), unfavorable channel conditions (e.g. enclosed and reverberant cavities), or performance requirements (e.g. ultra-low latency deadlines). Fortunately, these new environments may also offer optimization opportunities that are typically not exploitable in conventional scenarios. In any case, these aspects force designers to rethink the entire protocol stack \cite{jornet2014femtosecond, franques2021fuzzy} and adapt it to the specific conditions of these new and often extreme environments.

Among the new avenues to explore, this paper focuses on the application of wireless communications in the terahertz band within computing packages \cite{Sujay2012}. Such an idea is motivated by the recent emergence of massively parallel and heterogeneous computer architectures whose bottleneck generally lies in the intra-/inter-chip communications \cite{Shao2019}. In this context, communication has been traditionally served via on-chip networks formed by integrated routers linked with on-chip and chip-to-chip wired links, with simple topologies, routing and flow control mechanisms \cite{bertozzi2015fast}. However, as architectures scaled to handle massive parallelism and heterogeneity, these topologies suffered from multi-hop latency and energy efficiency issues, especially for collective communication patterns \cite{Karkar2016} and when moving from chip to chip \cite{Shamim2017}. 

Wireless communications at the chip scale, where on-chip antennas establish intra-/inter-chip links using the computing package as propagating medium, have become a promising solution to overcome the profound difficulties demonstrated by current interconnect solutions \cite{Sujay2012}. This is due to their inherent broadcast capabilities, low latency (even across chips), and flexibility to adapt to varying traffic patterns. In accordance to such promise, several groups have done extensive research aimed at achieving ultra-low latency and error rate with small area and energy footprints \cite{Pano2020a,Karkar2016,Shamim2017,Matolak2013CHANNEL,Rayess2017}. 

However, computing packages are inherently reverberant, hence creating lengthy delay spreads that hinder high-rate data transmissions with severe Inter-Symbol Interference (ISI) effects. Moreover, high energy spread prevents the creation of multiple spatial channels due to co-channel interference. These aspects have been generally neglected in most prior work \cite{timoneda2018channel}, although they are clearly incompatible with the high-speed and efficient vision for wireless communications at the chip scale. 

\begin{figure}[!t]
\centering
\includegraphics[scale=0.35]{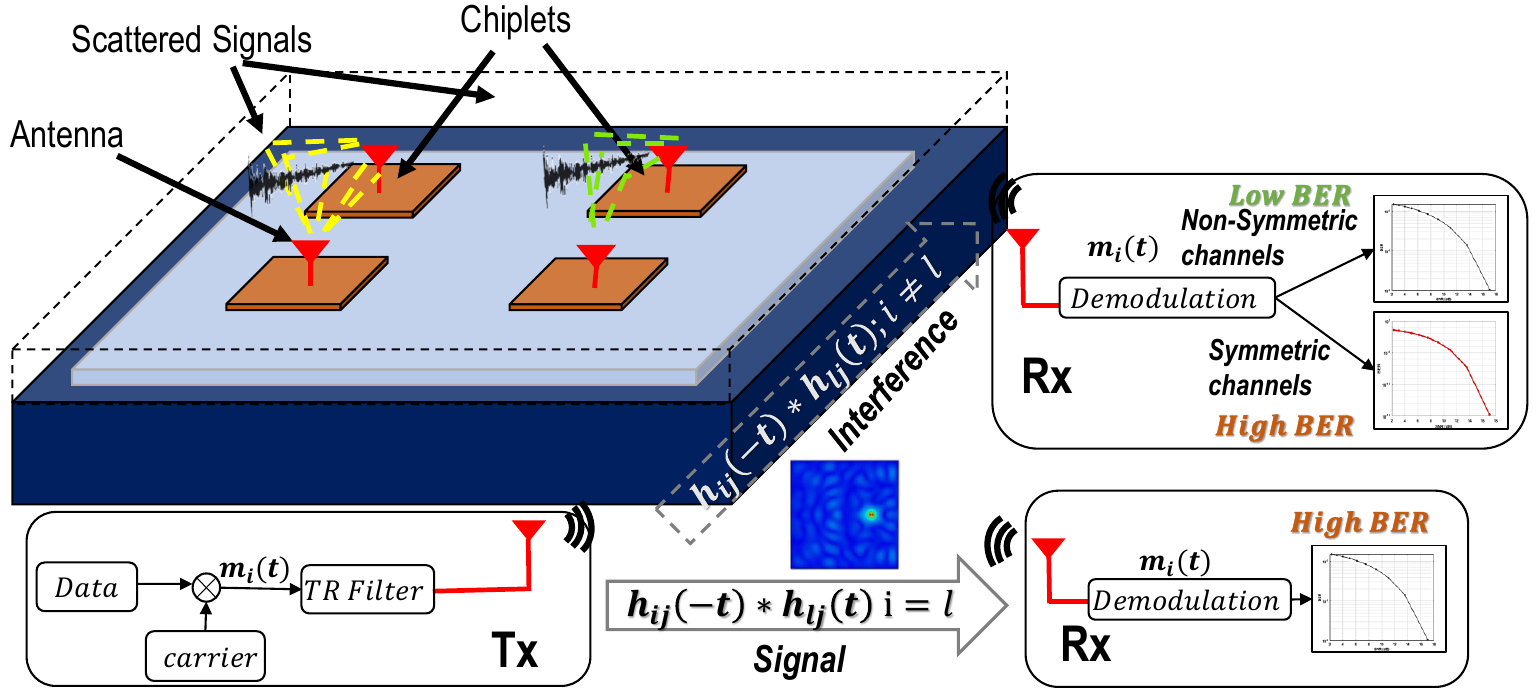}
\vspace{-0.1cm}
\caption{Time reversal for multi-channel wireless communications in a computing package. Through a filter at the transmitter end, time reversal attains spatiotemporal focusing at an intended receiver. If channels are not very highly correlated, multiple concurrent transmissions are possible.}
\label{fig:abstract}
\vspace{-0.4cm}
\end{figure}

In this paper, we aim to bridge this gap by jointly addressing the problems of inter-symbol and co-channel interference in reverberant chip packages. To this end, we propose to use Time Reversal (TR) \cite{Han2012, Xu2018, alexandropoulos2022time}. In essence, TR pre-emphasizes the transmitted symbols with the time-reversed version of the channel's response. This creates a spatial matched filter for an intended receiver or, in other words, focuses the signals in time and space around such receiver. As shown in Figure~\ref{fig:abstract}, the transmitter targets a specific receiver by choosing the appropriate TR filter. Then, as long as the impulse responses of two different links have a low correlation, multiple concurrent TR transmissions are possible in the same frequency band. This opens the door to the implementation of collective communication patterns that are widespread and extremely relevant in modern computing systems \cite{abadal2016characterization, Karkar2016}.  

While TR and other pre-coding alternatives like CDMA \cite{Vijayakumaran2014}, OFDMA \cite{Hakat2022} and zero-forcing beamforming \cite{Iwabuchi2023} may be difficult to adopt due to the complexity of the signal processing involved, the particularities of the on-chip scenario allow to simplify them significantly. In our case, TR requires knowing the channel response at all times. Fortunately, unlike in conventional scenarios, the wireless in-package channel is static and can be pre-characterized accurately.



In previous work, we explored the TR technique to enhance a single link within a single-chip architecture at 60 GHz \cite{bandara2023exploration}. In this paper, we instead propose a practical scheme to apply TR for multi-channel wireless communications within a multi-chip architecture (Figure \ref{fig:abstract}). In particular, we demonstrate how TR mitigates both ISI and co-channel interference in a frequency
on the lower side of the terahertz spectrum relevant to current efforts in 6G networks and beyond (i.e. 140 GHz).

After providing background on the chip-scale communications field in Section \ref{sec:background}, we model different multi-channel TR schemes according to Figure~\ref{fig:cases} in Section \ref{sec:model}. We simulate an actual computing package with realistic antennas at 140 GHz and demonstrate that the use of TR allows having multiple simultaneous transmissions that are robust against ISI, 
as described in Section \ref{sec:results}. Finally, we discuss feasibility and possible extensions of the work in Section \ref{sec:discussion} and conclude the paper in Section \ref{sec:conclusion}.


%% file: 2-background-short.tex
Next, we provide some background about the rationale and concept of Network-on-Chip (NoC) in Section \ref{sec:backNoc}, about the fundamentals of wireless chip-scale communications in Section \ref{sec:backWi}, about the wireless channels within computing packages in Section \ref{sec:backChan}, and about integrated antennas and their placement in Section \ref{sec:backAnt}.

\subsection{Network-on-Chip}
\label{sec:backNoc}
In the last decade, NoC has been the standard interconnect fabric inside chips \cite{Nychis2012}. NoCs are packet-switched networks of integrated routers and wires, typically arranged in a mesh topology \cite{Shao2019}. This approach, however, has important limitations when scaled beyond tens of processors. In particular, chip-wide and broadcast transfers suffer from increasing latency and energy consumption due to the many hops needed to reach the destination(s) \cite{Nychis2012, abadal2016characterization}. 

This problem cannot be fixed with longer interconnects due to the dramatic growth in resistivity of copper wires in deeply scaled technology nodes. Further, when moving to multi-chip processors, the concept of NoC has been extended to incorporate off-chip links and form the Network-in-Package (NiP) \cite{Shao2019}, aggravating their issues. For instance, NiPs are rigid and prone to over-provisioning. This, along with the limited amount of physical connection pins, clearly hinders the scalability of the wired approach.

\subsection{Wireless Chip-Scale Communications}
\label{sec:backWi}
Wireless communications at the chip scale have been proposed to comply with the demands of bandwidth and reconfigurability that NoCs fail to provide. Complementing NoCs with wireless links supplies broadcast support, reconfigurability, and low latency even between distant processors. 


Several efforts have investigated the impact of having wireless links at the network and system levels. It has been shown that complementing existing wired NoCs/NiPs with wireless channels for broadcast and long-range links \cite{Karkar2016, Shamim2017} can reduce the latency by an order of magnitude and improve throughput significantly. 
Further, it has been demonstrated that the computing speed of several CPU and accelerator architectures can be improved up to 10$\times$ \cite{franques2021widir}. For these impacts to be realized, however, wireless channels need to be shared among a large number of antennas. Hence, MAC protocols have been a crucial part of the wireless NoC research. Several variants of CSMA \cite{mestres2016mac} and token passing \cite{ franques2021fuzzy} uniquely tailored to the chip-scale environment have been proposed. 

At the physical layer, most wireless NoC proposals have advocated for low-order modulations that avoid power-hungry components, seeking to meet the stringent latency ($<$10 ns), speed ($>$100 Gb/s), error ($<$10\textsuperscript{-10}), energy ($<$1 pJ/bit) and area ($<$0.1 mm\textsuperscript{2}/Gb/s) requirements. As a result, On-Off Keying (OOK) prototypes at high frequencies have been presented in multiple works. 
In \cite{yu2014architecture}, a direct OOK modulator and non-coherent receiver is implemented with 65-nm CMOS, achieving 16 Gb/s at 60 GHz. The authors also present a triple-channel scheme achieving 48 Gb/s with 2 pJ/bit at an error rate below 10\textsuperscript{-12} and taking 0.75 mm\textsuperscript{2} of chip area. In \cite{Holloway2021,Yi2021}, the authors discuss how such designs could be taken to 300 GHz with SiGe and InP technology as well.

\subsection{Wireless Channels within Packages}
\label{sec:backChan}
Typically a chip consists of a lossy silicon substrate with several layers of metal stack on top and separated by dielectric layers \cite{markish2015chip}. There are several ways of connecting a chip with the rest of the system and enclosing it within the package. A commonly used package in modern computing systems is the flip-chip, for which multiple variants exist. 
The classical flip-chip package has, from the PCB up, a layer of solder bumps, followed by a thin layer of dielectric insulator, where the metallization layers are, with the silicon substrate and the heat sink on top \cite{wright2006characterization}.  
To interconnect multiple chips without going through the PCB, recently chips have started being placed side-to-side on top of a silicon interposer via an array of micro-bumps. The interposer is connected to the PCB through the solder bumps \cite{zhang2015heterogeneous}. For the sake of generality, the work developed in this paper takes place in an interposer package. 



Since the flip-chip package is largely enclosed, the chip propagation medium can be seen as a reverberation chamber \cite{Matolak2013CHANNEL}. All the energy propagating around the structure produces a multipath-rich environment that creates notches in the frequency response and lengthens the channel in time. The actual length of the channel depends mostly on the dimensions and resistivity of the silicon layer, which is a major contributor to the channel losses \cite{imani2021smart}. In any case, as we increase the symbol rate, the channel introduces significant ISI which undermines the quality of transmission \cite{Rayess2017}.

The channel in this scenario, though complex, is deterministic and practically invariant in time. The geometry, characteristics of the materials, and location of the nodes is decided beforehand and remains static after fabrication and integration. With this information, we can have an accurate pre-characterization that simplifies the task of compensating for the channel's impairments using techniques like TR, which is the case in this paper.

\subsection{Integrated Antennas}
\label{sec:backAnt}
The placement of an antenna within a computing package is complex as it is a densely integrated environment. In general, there are two strategies that we refer to on-chip integration and in-package integration. Several alternatives have been implemented across the spectrum, from millimeter-wave to optics, using both strategies \cite{Rayess2017, timoneda2018channel, bellanca2017integrated}. 

On the one hand, on-chip antennas can be horizontally or vertically co-integrated with the transceiver circuits. Horizontal designs generally use the metallization layers of a chip \cite{branch2005wireless} to implement variants of a dipole (meander, zig-zag), bow-tie or Yagi-Uda antenna \cite{gutierrez2009chip, Rayess2017, bellanca2017integrated}. The alternative to the horizontal approach is the implementation of antennas via a vertical approach, as in \cite{timoneda2018channel,Pano2020a}, where the authors propose to use Through-Silicon Vias (TSVs) as vertical monopoles with the metal layer acting as ground plane. This allows for the radiation to happen laterally towards other antennas. Although vertical antennas only support high frequencies due to the relatively small lengths of typical TSVs, in this paper we follow this approach due to its superior co-planar coupling and reduced area footprint.

On the other hand, in-package integration refers to the stacking of an antenna module (with its own ground plane and substrate) on top of the chip and the interconnection of the on-chip transceiver and the in-package antenna with a TSV. This allows to electromagnetically isolate the antenna from the layers underneath, reducing losses and allowing for a wider range of antenna designs, at the cost of adding manufacturing steps, cost, and reverberation \cite{yeh2012design}. 




%% file: 3-model.tex

\begin{figure}[!t]
\centering
\includegraphics[width=\columnwidth]{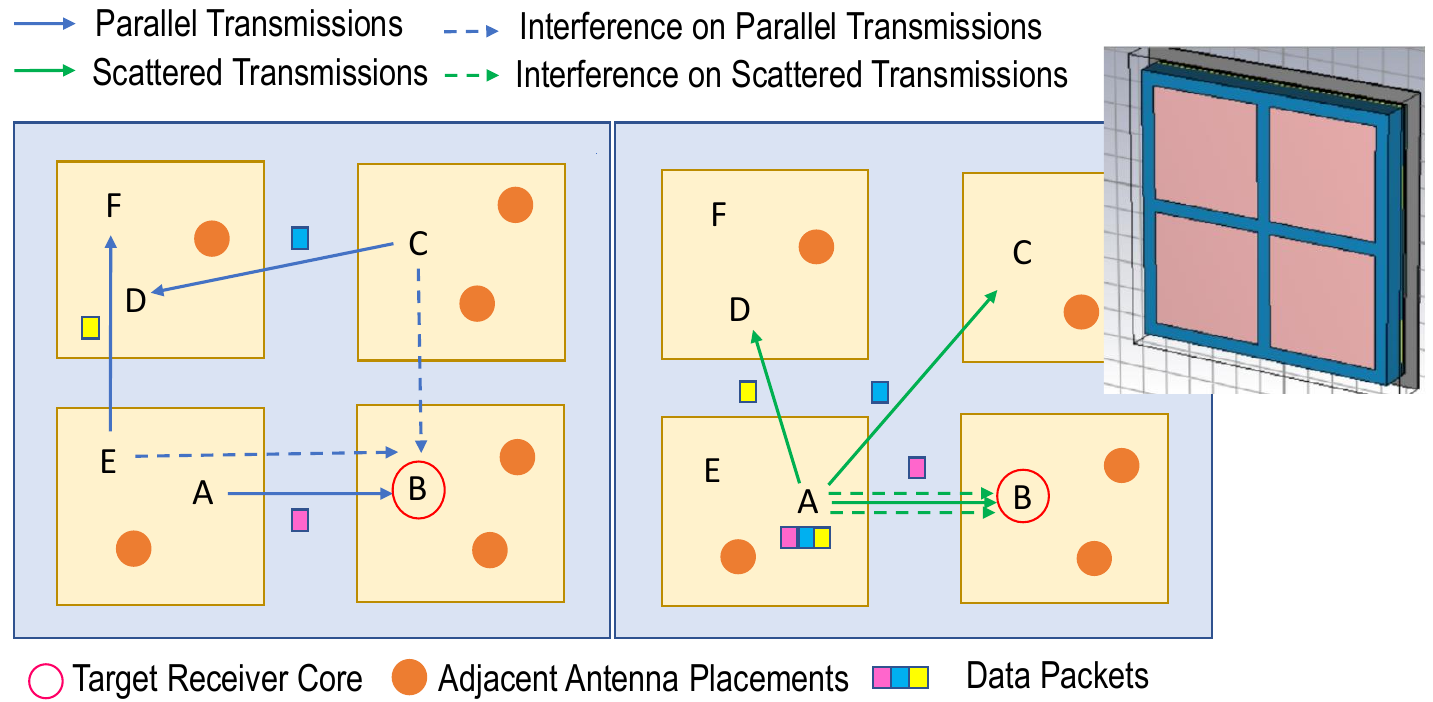}
\vspace{-0.5cm}
\caption{Evaluated multi-receiver time-reversal scenarios, with multiple concurrent transmitters (left) or a single transmitter sending, at the same time, different data to different receivers (right). The inset shows the simulated package in CST.}
\label{fig:cases}
\vspace{-0.4cm}
\end{figure}

In this section, we model the three different cases explored in this paper as shown in Figure \ref{fig:cases}. 


\subsection{Single TR Link}
Let us first assume a single link from an arbitrary node $i$ to an arbitrary node $j$. With prior characterization of the CIR for the $i\rightarrow j$ link, we can assume $h_{ij}(t)$ known by the transmitter. 

On the one hand, the symbols modulated by node $i$ will be pre-coded with the time reversal filter, leading to
\begin{equation}
x_{i}(t) = h_{ij}(-t) * m_{ij}(t),
\label{eqn:trnodeA}
\end{equation}
where $x_{i}(t)$ are precoded symbols to be transmitted from node $i$ to node $j$, $*$ is the convolution operation, and $h_{ij}(-t)$ is the time-reversed CIR of the link, which can be expressed as
\begin{equation}
h_{ij}(t) = \sum_{k}\alpha_{k} e^{j\psi_{k}}\delta(t-\tau_{k}),
\label{eqn:cnannel}
\end{equation}
with $\{\alpha_k, \psi_{k}, \tau_k \}$ being the amplitude, phase and propagation delay of the multi-path component $k$ of the channel $i\rightarrow j$. 

On the other hand, the signal received at node $j$ placed in an arbitrary position can be expressed, in the time domain, as
\begin{equation}
y_{j}(t) = h_{ij}(t) * h_{ij}(-t) * m_{ij}(t) + n_{j}(t) 
\label{eqn:Rx_enduser}
\end{equation}
where $n_{j}(t)$ is the noise at receiver $j$, assumed of zero mean and variance of $\sigma^2$ with power $N = k_{B}T B$ with $k_{B}$, $T$, and $B$ being the Boltzmann constant, temperature, and bandwidth. 



As can be observed, Eq.~\eqref{eqn:Rx_enduser} is similar to the usual matched filter expression when the pre-coding filter matches with the link's CIR. As we will see, this produces a peak in the received signal. For any other receiver $l \neq j$, the pre-coding filter convoluted by the $i\rightarrow l$ link CIR leads to a poor signal strength at the receiver, unless both channels are highly correlated, which could happen due to spatial symmetries.


\subsection{Multiple Transmitters}\label{multlink}
Following up from the case above, let us now consider that $L$ TR links are being deployed concurrently and analyze the resulting signal received in $j$. By the superposition principle, we have that the received signal is the contribution of all simultaneous transmissions, this is
\begin{equation}
y_{j}(t) = n_{j}(t) + \sum_{l\in L} h_{lj}(t) * h_{l}(-t) * m_{l}(t),
\label{eqn:Rx_enduserI}
\end{equation}
where $h_{l}(-t)$ and $m_{l}(t)$ are the time-reversal filter and modulated stream of transmitter $l$ to target its intended destination. This can be also expressed as:
\begin{equation}
\begin{split}
y_{j}(t) = h_{ij}(t) * h_{ij}(-t) * m_{i}(t) + n_{j}(t)  + \\ + \sum_{l\in L,l\neq i} h_{lj}(t) * h_{l}(-t) * m_{l}(t).
\end{split}
\label{eqn:Rx_enduserI2}
\end{equation}
or, in terms of spatial correlation,
\begin{equation}
y_{j}(t) = \underbrace{R_{ij}(t) *  m_{i}(t)}_\text{signal} + \sum_{l\in L,l\neq i} \underbrace{R^{'}_{lj}(t) * m_{l}(t)}_\text{interference} + \underbrace{n_{j}(t)}_\text{noise},
\label{eqn:Rx_enduserI2corr}
\end{equation}
where $R_{ij}(t)$ and $\sum_{l\in L,l\neq i} R^{'}_{lj}(t)$ represent the auto-correlation and the sum of cross-correlation functions in spatial domain for multiple transmitters, respectively. The equation clearly differentiates between signal, noise, and interference. 
The interference will be proportional to the number of interferers and also to the correlation between the different CIRs, as discussed above

\subsection{Multiple Receivers}
Next, we analyze the case where a single transmitter is encoding $S$ modulated streams using different pre-coding filters, each targeting a specific receiver \cite{Han2012}. This is equivalent to the collective pattern \textsc{MPI\_Scatter}, extremely relevant in computer architecture. The resulting signal is
\begin{equation}
x_{i}(t) = \sum_{s\in S} h_{is}(-t) * m_{is}(t),
\label{eqn:trnodeAmulti}
\end{equation}
where $m_{is}$ is the modulated stream from node $i$ targeted to node $s$ and $h_{is}(-t)$ is the time-reversal filter for the path $i\rightarrow s$. Hence, the radiated waveform contains components from all the different streams directed to all $S$ receivers. On the other side, the signal at an arbitrary receiving node $j$ is
\begin{equation}
y_{j}(t) = n_{j}(t) + h_{ij}(t) * \sum_{s\in S} h_{is}(-t) * m_{is}(t),
\label{eqn:Rx_enduserImulti}
\end{equation}
which is the convolution of the transmitted signal with the channel $i\rightarrow j$ plus thermal noise. Expanding the sum yields 
\begin{equation}
\begin{split}
y_{j}(t) = h_{ij}(t) * h_{ij}(-t) * m_{ij}(t) + n_{j}(t)  + \\ + \sum_{s\in S,s\neq j} h_{ij}(t) * h_{is}(-t) * m_{is}(t).
\end{split}
\label{eqn:Rx_enduserI2multi}
\end{equation}

\noindent Again, Eq. \eqref{eqn:Rx_enduserI2multi} can be rewritten with spacial correlation as
\begin{equation}
y_{j}(t) = \underbrace{R_{ij}(t) *  m_{ij}(t)}_\text{signal}+\sum_{s\in S,s\neq j} \underbrace{R^{'}_{sj}(t) * m_{is}(t)}_\text{interference}+\underbrace{n_{j}(t)}_\text{noise},
\label{eqn:Rx_enduserI2multi1}
\end{equation}
where $\sum_{s\in S,s\neq j} R^{'}_{sj}(t)$ represents the sum of cross-correlation functions in spatial domain for multiple receivers. The signal term should be maximized by virtue of the spatiotemporal TR effect, whereas interference will be minimized as long as the channels between $i$ and the participating receivers are loosely correlated.

%% file: 4-results-short.tex
To evaluate the proposed approach, we first simulate the chip environment using CST Microwave Studio \cite{CST}. We build a realistic model of an interposer package described in Sec.~\ref{sec:backChan} with twelve antennas in four chiplets (Fig. \ref{fig:cases}). Chiplets are 5$\times$5 mm\textsuperscript{2} with 0.5-mm high-resistivity (HR) silicon and 0.5-mm aluminum nitride as heat spreader, on top of a 0.1-mm HR silicon interposer. We obtain the CIR by means of the full-wave solver and import it to MATLAB, where we apply the system model and simulate the transmission of 1000 bits with ASK modulation. Our aim is to assess the possibility of having concurrent transmissions with distinct TX-RX pairs (Sec. \ref{mtr}) and from one TX to multiple RX (Sec. \ref{mrx}).


\begin{figure*}[!t]
\centering
\begin{subfigure}[t]{0.22\textwidth} 
\includegraphics[width=\textwidth]{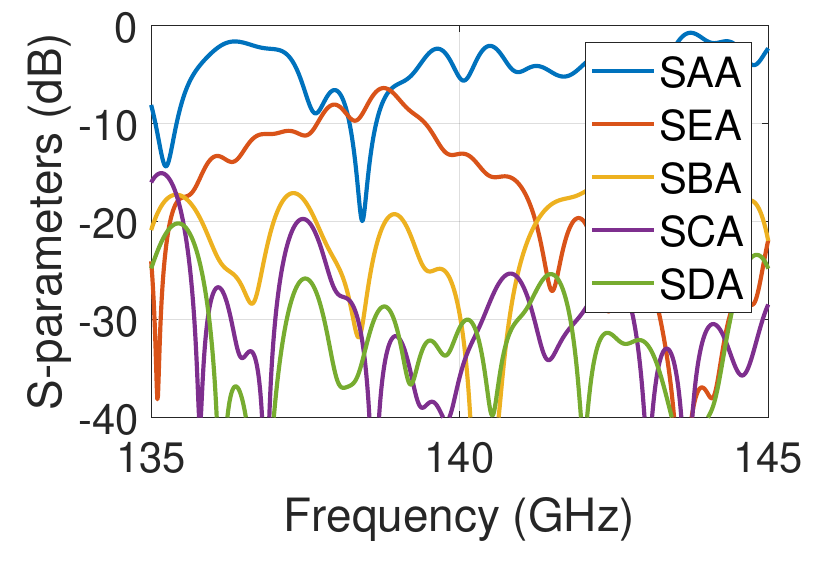}
\caption{}
\end{subfigure}
\begin{subfigure}[t]{0.34\textwidth}
\includegraphics[width=0.47\textwidth]{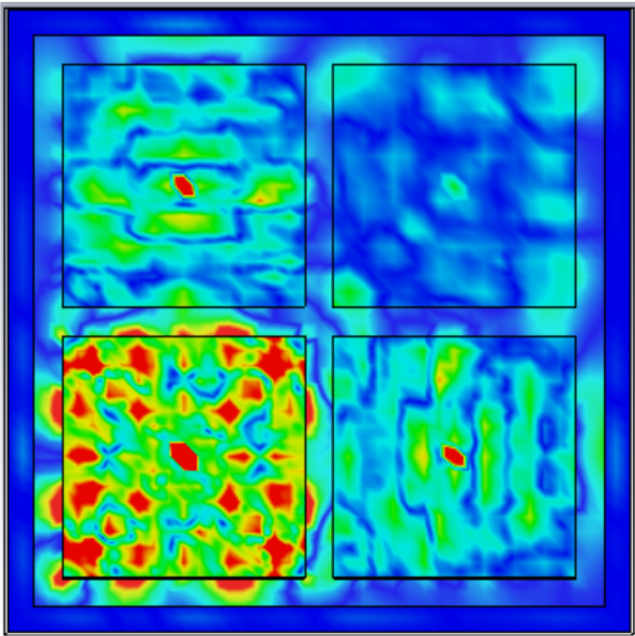} 
\includegraphics[width=0.47\textwidth]{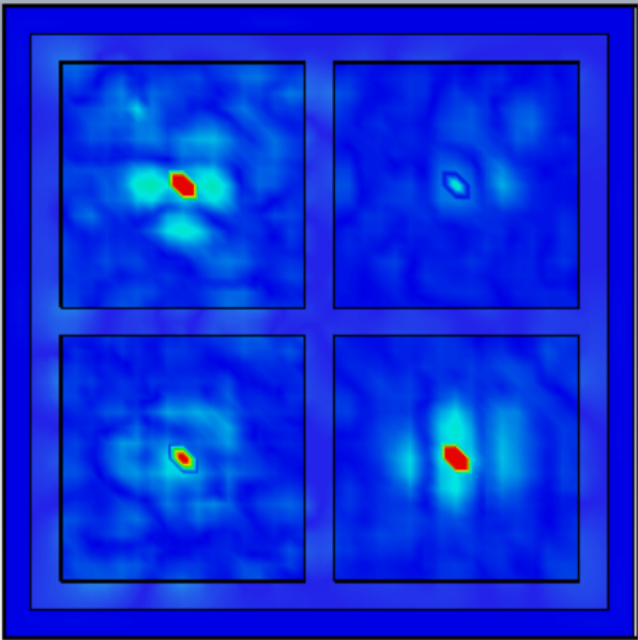}
\caption{}
\end{subfigure}
\begin{subfigure}[t]{0.22\textwidth}
\includegraphics[width=\textwidth]{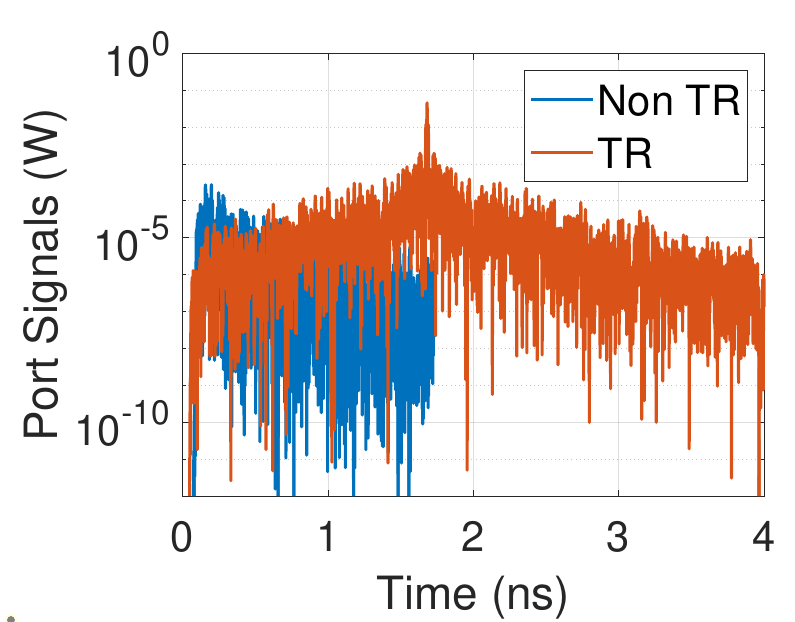}
\caption{}
\end{subfigure}
\begin{subfigure}[t]{0.2\textwidth}
\includegraphics[width=\textwidth]{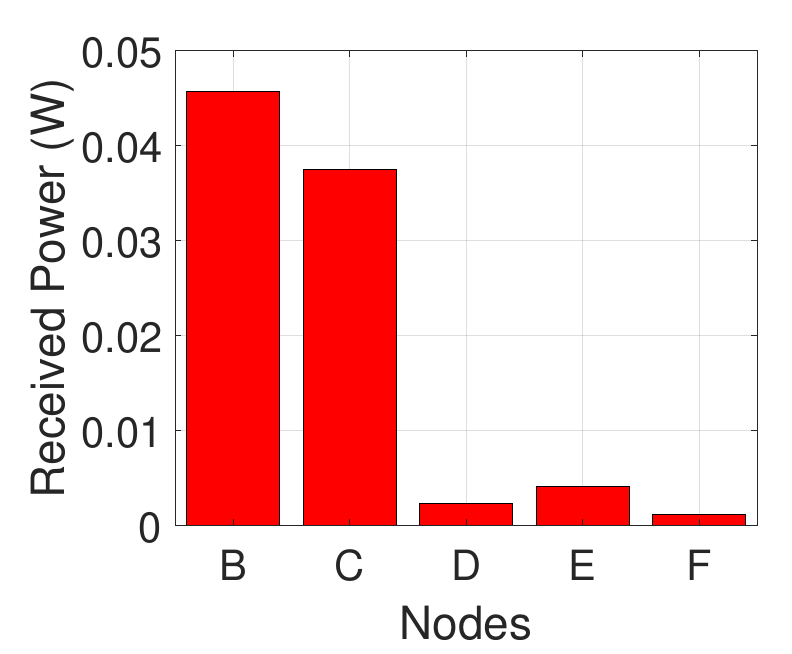}
\caption{}
\end{subfigure}
\vspace{-0.3cm}
\caption{Single-link time-reversal with vertical monopoles in an interposer package at 140 GHz. (a) S-parameters of the transmitting antenna and all links. (b) Spatial distribution of field before and applying TR. (c) Signal received over time in node $B$ before and after applying TR. (d) Total received power in the intended receiver $B$ and non-intended receivers $C$--$F$.} 
\label{fig:channel}
\vspace{-0.4cm}
\end{figure*}

\subsection{Wireless Channel}
Figure \ref{fig:channel} summarizes the simulations made to illustrate how TR works in an interposer package with vertical monopoles operating at 140 GHz. Fig.~\ref{fig:channel}(a) shows how the antennas resonate at $\sim$138 GHz. The power received across the 135--145 GHz band shows notches, suggesting the presence of strong multipath. Panels \ref{fig:channel}(b) and \ref{fig:channel}(c) demonstrate the spatiotemporal focusing effect of TR by plotting the spatial and temporal field distribution, respectively. Spatially, we observe the non-TR transmission being scattered across the four chiplets, while the TR case is clean and only focuses the energy around very specific spots. 
In time, the maximum instantaneous power concentration for a non-TR transmission is 0.2 mW, far from the 45 mW achieved in the TR case. Finally, Fig.~\ref{fig:channel}(d) shows how TR compresses the interference to non-targeted receivers, as long as their channel is not correlated with the intended one. Here, this is the case of node $C$, which is located in a close-to-symmetric position to $B$ with respect to $A$.



\begin{figure}[!t]
\centering
\begin{subfigure}[t]{0.46\columnwidth}
\includegraphics[width=\textwidth]{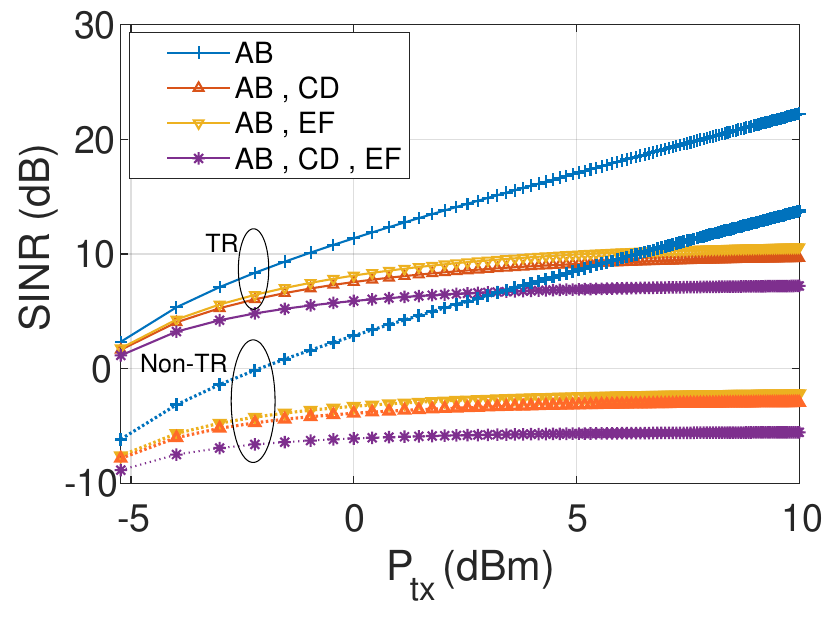}
\caption{}
\label{fig:multi-linka}
\end{subfigure}
\hspace{-0.4cm}
\begin{subfigure}[t]{0.46\columnwidth}
\includegraphics[width=\textwidth]{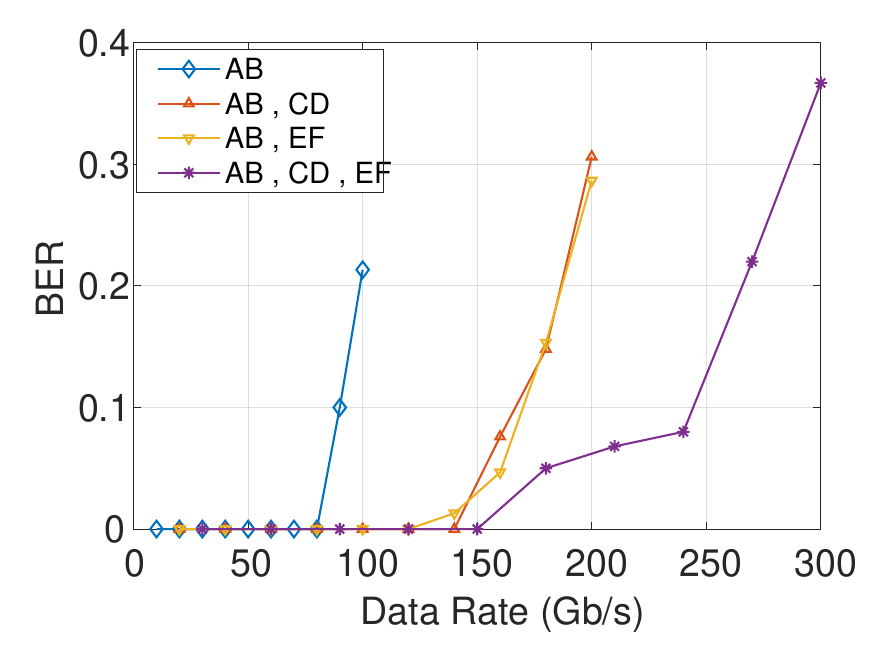}
\caption{}
\label{fig:multi-linkc}
\end{subfigure}
\vspace{-0.2cm}
\caption{Time reversal with multiple concurrent transmitters. (a) 
SINR for non-TR and TR cases with single or multiple links as a function of the transmitted power at 50 Gb/s. 
(b) BER as a function of the aggregated data rate for a power of 10 dBm per transmitter in the TR case.}
\vspace{-0.4cm}
\label{fig:multi-link}
\end{figure}

\subsection{Multiple Transmitters}\label{mtr} 
Next, as shown in the left panel of Fig.~\ref{fig:cases}, we excite node $A$ with the time-reversed CIR of the $AB$ path, and simultaneously do the same for the $CD$ and $EF$ links. To evaluate the capacity of TR to support multiple channels, we obtain the SINR and BER at the receivers for $AB$ alone, $AB$ with $CD$ or $EF$, and the three transmissions concurrently.







When sweeping the transmission power, we observe in Figure \ref{fig:multi-linka} how there is an 8 dB SINR gap between TR and non-TR transmissions, which confirms that TR minimizes ISI. With two concurrent transmissions, the interference degrades the SINR to the point of being dominant at high power conditions. Moreover, a $\sim$1 dB difference is observed between the interference caused by $CD$ or $EF$. As expected, the effect of both interferent links accumulates.


In Figure \ref{fig:multi-linkc} the BER was measured as a function of the aggregated data rate with a fixed transmission power of 10 dBm per transmitter. As expected, the increase of data rate reduces the energy per bit and, hence, increases the BER. For the simulated single-link scenario, transmissions over 80 Gb/s become impractical as the BER scales quickly to unacceptable levels. In comparison, the BER for the non-TR single-link transmission saturates below 10 Gb/s (not shown for the sake of clarity). We also observe that the two-link and three-link cases achieve a higher total data rate, since the interference is not strong enough to impair the individual links.


\subsection{Multiple Receivers} \label{mrx}
Next, we show the results of transmitting from $A$ to $B$-$D$. We remind that $A$ sends different chunks of data to each receiver thanks to TR multiple access as described in \cite{Han2012}. \par

\begin{figure}[!t]
\centering
\begin{subfigure}[t]{0.52\columnwidth}
\includegraphics[width=\textwidth]{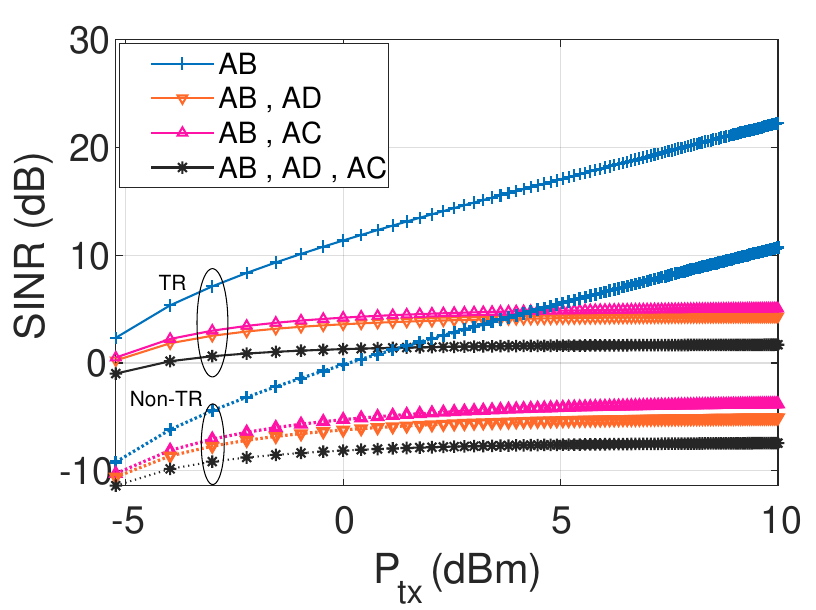}
\caption{}
\label{fig:scattera}
\end{subfigure}
\hspace{-0.4cm}
\begin{subfigure}[t]{0.46\columnwidth}
\includegraphics[width=\textwidth]{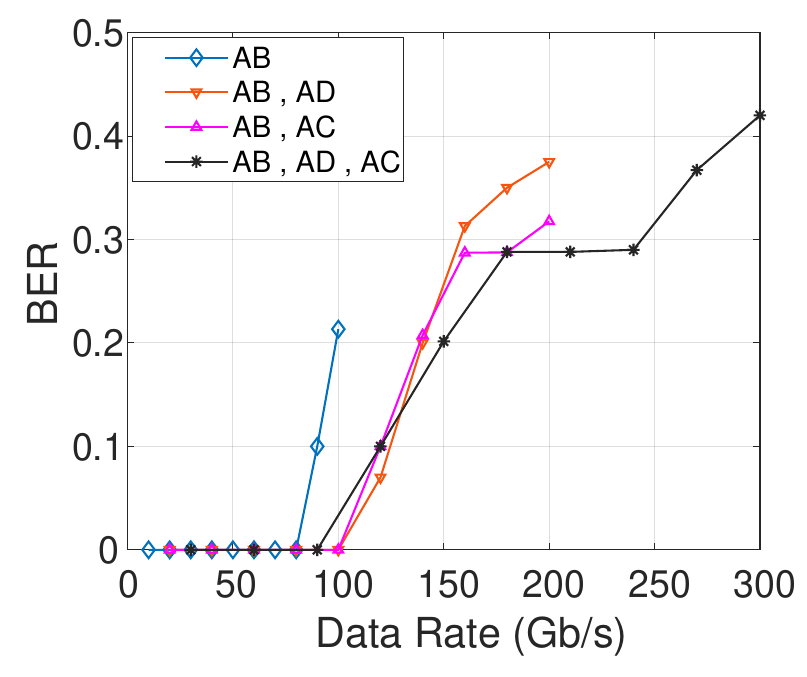}
\caption{}
\label{fig:scatterc}
\end{subfigure}
\vspace{-0.2cm}
\caption{Time reversal with multiple concurrent receivers. (a) 
SINR for non-TR and TR cases with single or multiple receivers and increasing the transmitted power, at 50 Gb/s. 
(b) BER as a function of the aggregated data rate for a total power of 10 dBm in the TR case.}
\vspace{-0.4cm}
\label{fig:scatter}
\end{figure}

As similar to in Section \ref{mtr}, the SINR and BER was analysed in receiver $B$, with the impact of co-channel interference on concurrent transmissions. As observed in Figure \ref{fig:scattera}, the SINR for multiple transmissions have been decreased in compared to the measurements in Figure \ref{fig:multi-linka}, as in single transmitter multiple TR links are active to perform concurrent transmissions. With the effect of multi-path components of CIR in superimposed channels, the focused EM distribution could reduce the received signal strength with a minor amount, while maintaining the interference suppression.


The BER measurements in Figure \ref{fig:scatterc}, shows the effect of interference on parallel receivers for 10 dBm transmit power as we increase data rate. A similar BER performance with increased interference is observed as compared with Figure \ref{fig:multi-linkc}, while the BER take higher
values with minor increment based on superimposed TR transmissions.

%% file: 5-discussion-short.tex
\noindent \textbf{Multicast.} TR can be used to achieve a significant concentration of energy in time and space and, hence, simultaneously compress inter-symbol and co-channel interference. However, the broadcast appeal of wireless is lost. Yet if we consider a scenario with many antennas carefully placed, we can use symmetries to have multiple positions where the terminals will obtain the same response. To illustrate this, let us refer to Figure \ref{fig:channel}(b), where we see two peaks of highly concentrated field. These peaks are obtained at unison with a single excitation of the transmitter because they are placed in a symmetric location that makes their channel responses identical. 

\noindent \textbf{Interference Hotspot.} As a downside, symmetries can also lead to undesirable effects in multi-link TR schemes, as shown in Figure~\ref{fig:channel}(d). In that case, receiver $C$ is in a position which is not spatially symmetrical to $B$, but that it shows a close resemblance with it in terms of CIR. Hence, in highly correlated spots, concurrent transmissions are not advisable. 


\noindent \textbf{Feasibility.} The TR technique has been evaluated here with infinite sampling resolution. However, real wireless NoC will need to resort to simpler filters to realize the TR potential. We will explore the trade-offs and limitations that arise with the use of non-ideal filters in future work.

%% file: 6-conclusion-short.tex
In this paper, we have proposed and evaluated the use of time-reversal (TR) to combat inter-symbol and co-channel interference in wireless communications inside and across chips of a computing package. 
We simulated an interposer package at 140 GHz for multiple scenarios with a single or multiple concurrent TR links.
Comparing the SINR of TR and non-TR transmissions, we have a heightened signal that shows TR improvement. With the increment of simultaneous wireless links, we got a decrement in SINR due to the correlation between links in positions close to symmetry. Similarly, when we increase the number of antennas receiving from a single sender, the SINR starts to decrease because there is a limit on the amount of CIR you can superimpose in a single transmission. We could check that the BER saturates faster with the increment of parallel transmitters and receivers. 
We have shown we can have parallel terminals transmitting in unison and multicast transmission in one-to-many fashion. All of these can be attained in the same frequency thanks to  the interference compression of TR.